\documentclass[aps,twocolumn,superscriptaddress,nofootinbib,a4paper,longbibliography]{revtex4-2}

\usepackage{times}
\usepackage{epsfig}
\usepackage{amsfonts}
\usepackage{amsmath}
\usepackage{esint}
\usepackage{amsthm}
\usepackage{amssymb}					
\usepackage{amsthm}
\usepackage{dsfont}
\usepackage{bm}
\usepackage{mathtools}
\usepackage{color}
\usepackage{multirow}
\usepackage{booktabs}
\usepackage{tabularx}
\usepackage[normalem]{ulem}
\newcommand{\stkout}[1]{\ifmmode\text{\sout{\ensuremath{#1}}}\else\sout{#1}\fi}
\usepackage{latexsym}
\usepackage{mathrsfs}
\usepackage{natbib}
\usepackage{verbatim}
\usepackage[T1]{fontenc}
\usepackage{float}
\usepackage{graphicx}
\usepackage{subcaption}
\usepackage{xcolor}
\usepackage{physics}
\usepackage{soul}
\usepackage{caption}
\usepackage{subcaption}
\usepackage{enumerate}
\usepackage[font=small,labelfont=bf]{caption}

\usepackage{microtype}

\usepackage{xspace}

\theoremstyle{definition}

\usepackage{amssymb}
\usepackage{amsthm}
\usepackage{mathtools} 
\usepackage[customcolors]{hf-tikz} 
\usetikzlibrary{patterns}
\usetikzlibrary{matrix,decorations.pathreplacing}

\pgfkeys{tikz/mymatrixenv/.style={decoration={brace},every left delimiter/.style={xshift=8pt},every right delimiter/.style={xshift=-8pt}}}
\pgfkeys{tikz/mymatrix/.style={matrix of math nodes,nodes in empty cells,left delimiter={[},right delimiter={]},inner sep=1pt,outer sep=1.5pt,column sep=2pt,row sep=2pt,nodes={minimum width=20pt,minimum height=10pt,anchor=center,inner sep=0pt,outer sep=0pt}}}
\pgfkeys{tikz/mymatrixbrace/.style={decorate,thick}}

\usepackage[colorlinks=true,linkcolor=blue,citecolor=magenta,urlcolor=blue]{hyperref}

\begin{document}
	

\title{Unlimited quantum correlation advantage from bound entanglement}

\author{Armin Tavakoli} \email{armin.tavakoli@fysik.lu.se}
\affiliation{Physics Department and NanoLund, Lund University, Box 118, 22100 Lund, Sweden}
\author{Carles Roch i Carceller}
\affiliation{Physics Department and NanoLund, Lund University, Box 118, 22100 Lund, Sweden}
\author{Lucas Tendick}
\affiliation{Inria, Université Paris-Saclay Palaiseau, France}
\affiliation{CPHT, CNRS, Ecole Polytechnique, Institut Polytechnique de Paris, Palaiseau, France}
\affiliation{LIX, CNRS, Ecole Polytechnique, Institut Polytechnique de Paris, Palaiseau, France}
\author{Tam\'as V\'ertesi}
\affiliation{HUN-REN Institute for Nuclear Research, P.O. Box 51, H-4001 Debrecen, Hungary}

\begin{abstract}
Entangled states that cannot be distilled to maximal entanglement are called bound entangled and they are  often viewed as  too weak to break the limitations of classical models. Here, we show a strongly contrasting result: that bound entangled states, when deployed as resources between two senders who communicate with a receiver, can generate correlation advantages of unlimited magnitude. The proof is based on using many copies of a bound entangled state to assist quantum communication. We show that in order to simulate the correlations predicted by bound entanglement, one requires in the many-copy limit either an entanglement visibility that tends to zero or a diverging amount of overhead communication. This capability of bound entanglement is unlocked  by only using elementary single-qubit operations. The result shows that bound entanglement can be a scalable resource for breaking the limitations of physical models without access to entanglement. 
\end{abstract}

\date{\today}
\maketitle


\section{Introduction} 
\begin{figure}[t!]
	\centering
	\includegraphics[width=0.9\columnwidth]{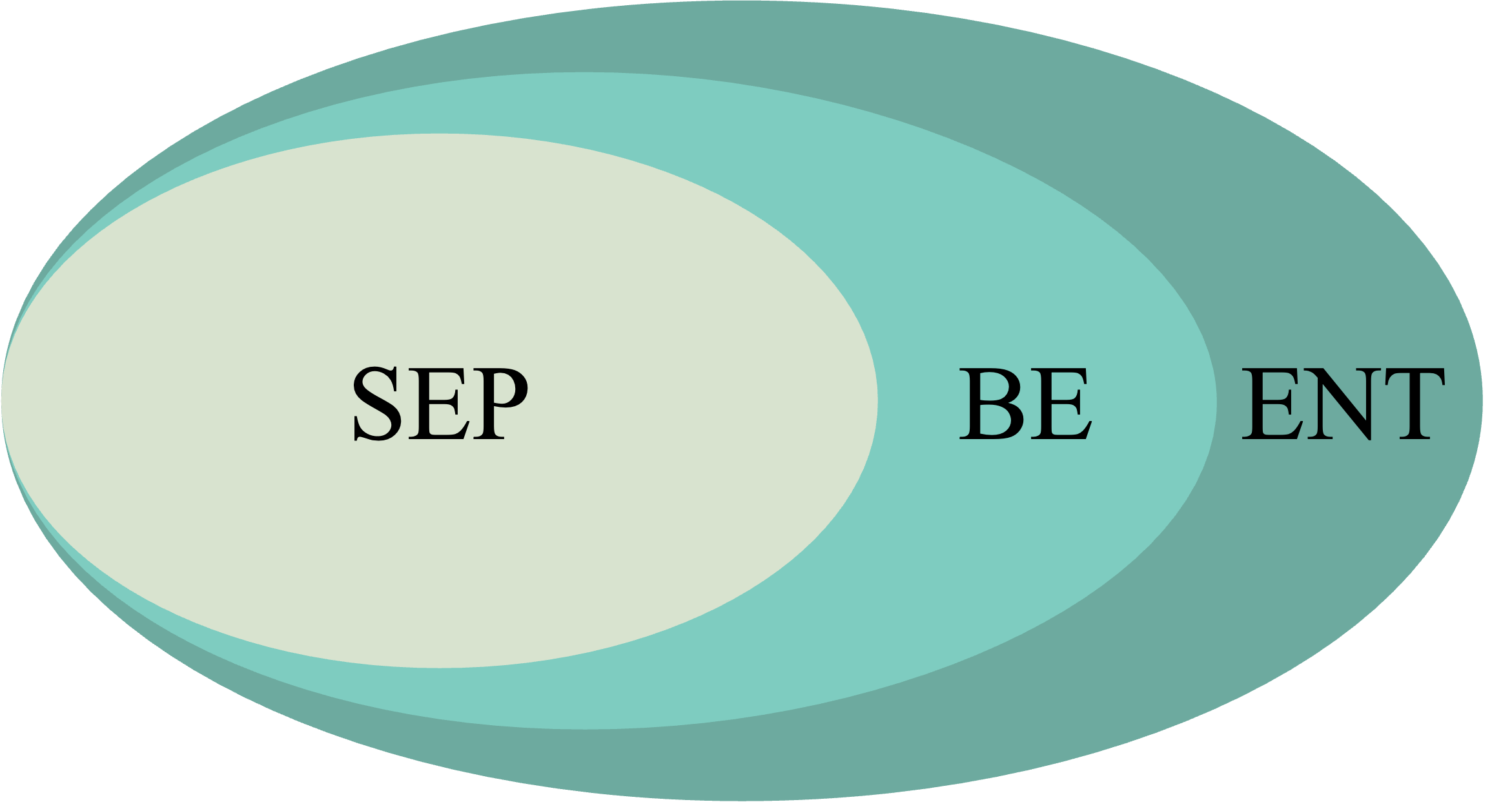}
	\caption{\textit{Bipartite quantum states.} The sets of separable states, bound entangled states and entangled states. All states that are not separable are  entangled. The subset of entangled states that are bound entangled cannot be distilled to maximal entanglement when available in any number of copies.}\label{Fig_BE}
\end{figure}

Entanglement is the paradigmatic resource for quantum information. Although most useful forms of entanglement can be converted into maximally entangled states by the process of entanglement distillation, there exist also entangled states for which distillation is impossible, even when the state is available in infinitely many copies. These states are called bound entangled \cite{Horodecki1998}; see Fig~\ref{Fig_BE}. In practice, bound entangled states are classified as the entangled states that fail to be detected by the seminal positive partial-transpose (PPT) criterion, as the existence of other bound entangled states remains a long-standing open problem \cite{IQOQIPProblem2, FiveOpenProblems}. Bound entangled states are remarkably common in high-dimensional Hilbert spaces: their volume becomes superexponentially larger than that of separable states \cite{Auburn2006}, and their distance to the separable set can become arbitrarily large \cite{Beigi2010}. \\
\indent Nevertheless, bound entangled states are often too weakly entangled to be useful for quantum information protocols, let alone to generate correlations that break the limitations of generic classical models (see the review \cite{Hiesmayr2024}). For example, bound entangled states can neither be used to beat the classical capacity limit in quantum communication \cite{Horodecki2001} nor the classical fidelity limit in teleportation \cite{Horodecki1999}
. Moreover, they have been conjectured to be useless for device-independent quantum key distribution \cite{Friedman2021}. For 15 years, the conjecture stood that bound entanglement also cannot violate any Bell inequality \cite{Peres1999}, before being falsified in 2014 by an explicit counter example \cite{Vertesi2014}. However, the degree of violation, which is of the order $10^{-4}$, is too small to make it practically relevant. Despite later works exploring nonlocality from bipartite bound entangled states, no significantly larger violation has been found \cite{Yu2017, Pal2017}. Similar trends have also been observed in quantum steering scenarios \cite{Moroder2014}. This may suggest that it is a rare occurance that a bound entangled state can violate, even by a tiny margin, a Bell inequality or steering inequality. \\
\indent In contrast, it was recently shown that three-dimensional bound entangled states can generate significant correlation advantages in scenarios that involve communication between senders and receivers \cite{Carceller2025}. Subsequently, numerical evidence has shown that the correlation advantages can become even more pronounced in communication tasks using four-dimensional bound entanglement \cite{Marton2025}. These results pave the way for a much stronger and conceptually important question:

\begin{center}
\textit{Can high-dimensional bound entanglement generate unlimited correlation advantages over any possible communication model that has no access to entanglement?} 
\end{center}

\noindent This question is the focus of this article, and we answer it affirmatively.

We consider a scenario with two independent senders and one receiver \cite{Bakhshinezhad2024}. The formers respectively hold private classical data and encode it   into quantum messages that are sent to the receiver. The receiver decodes the two messages with the aim of learning some binary property of their combined data. This is illustrated in Fig~\ref{Fig_scenario}. Importantly, no limitations are assumed on how the messages are encoded and decoded, apart from the capacity of the communication channel being restricted to $D$-dimensional systems.  In order to reveal the diverging correlation advantages that bound entanglement can generate in this scenario, we first consider the communication task numerically studied in Ref~\cite{Marton2025}. We prove their main conjecture,  namely that four-dimensional bound entanglement generates a sizable correlation advantage. Importantly, we also find that these advantages are common among bound entangled states: any such state that is witnessed through a specific well-known entanglement criterion can be used to generate a correlation advantage. Equipped with this result, we construct an $N$-fold parallel repetition of this task and determine the optimal performance possible without entanglement. Then, we show that by using  $N$ copies of a bound entangled state, one achieves an advantage whose magnitude diverges in the number of repetitions. This leads to scalable violations of correlation inequalities satisfied by all quantum models without entanglement.

We benchmark the advantage of bound entanglement in two distinct ways. Firstly, our parallel repetition protocol is based on the ability of the senders to transmit messages of dimension $4^N$. We show that the correlations predicted by this protocol cannot be reproduced without entanglement with any amount of communication below $6^N$. Consequently, in the limit of large $N$, the simulation of the correlation advantages  requires an unbounded amount of overhead communication, of either classical or quantum nature, which scales at least as $O(N)$ qubits. Secondly, we consider the amount of white (isotropic) noise that must be added to our $N$-copy bound entangled state in order to prevent a violation of the correlation inequalities. For this, we show that the visibility of the state must scale at least as  $O\left(0.667^N\right)$, which tends to zero in the many-copy limit. We conclude with a discussion on the potential capabilities of bound entanglement in breaking classical constraints and enhancing quantum information processing. \\

\section{Scenario}
\begin{figure}
	\centering
	\includegraphics[width=\columnwidth]{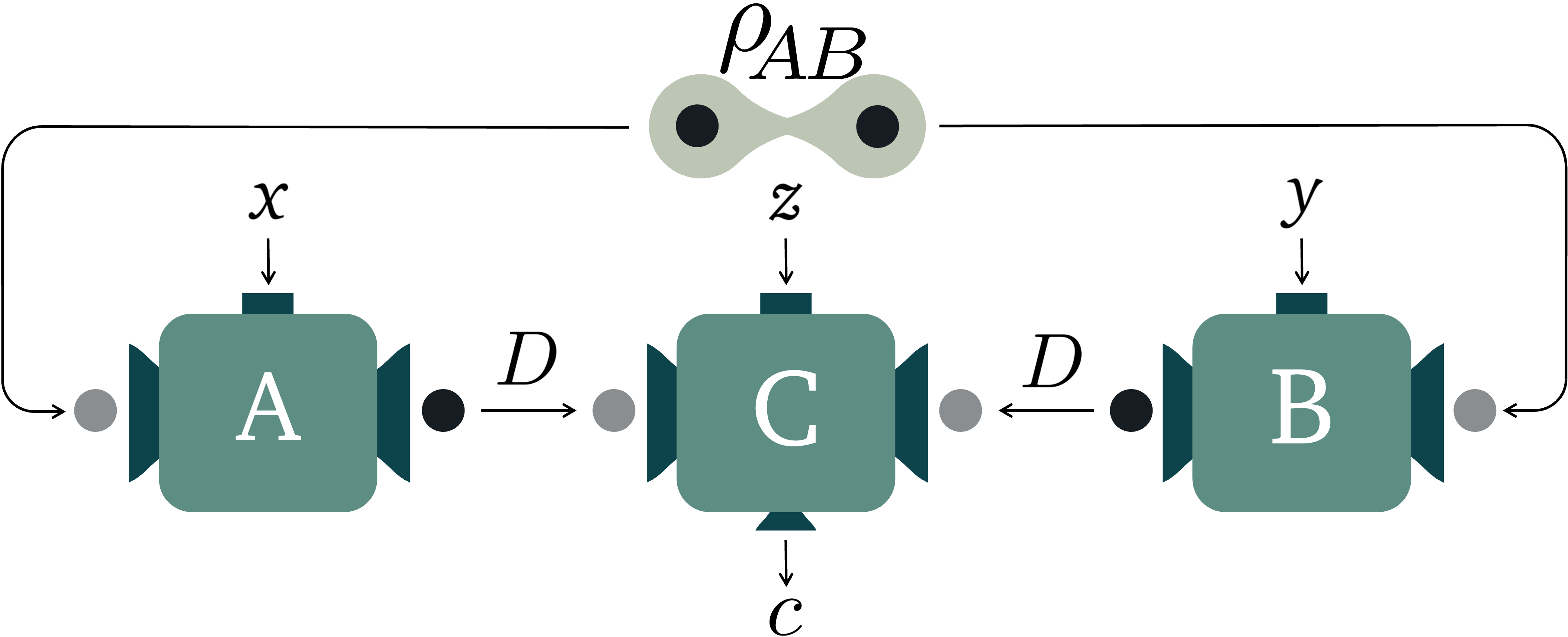}
	\caption{\textit{Communication scenario.} Alice and Bob encode their data $x$ and $y$ into $D$-dimensional messages that are sent to Charlie. Charlie selects a question, $z$, and produces an answer $c$. Alice and Bob can use a shared resource $\rho_{AB}$ to assist their encoding operations.}\label{Fig_scenario}
\end{figure}

Consider the scenario illustrated in Fig.~\ref{Fig_scenario}. The senders, named Alice and Bob, are allowed to share a bipartite state called $\rho_{AB}$. They privately select inputs $x\in[16]$ and $y\in[16]$ respectively and each encode their data into a quantum system of dimension $D$, denoted $\tau^A_x$ and $\tau^B_y$, respectively. Here, we define $[n] \equiv \{1,\ldots, n\}$. These encoding operations act on their respective halves of $\rho_{AB}$ and the resulting quantum messages are sent to the receiver called Charlie. Charlie selects an input $z\in[16]$, which corresponds to a binary function $w_z(x,y)\in\{+1,-1\}$. By measuring the incoming states with an observable $C_z$ he outputs $c\in\{+1,-1\}$ as his guess for the function value. To define the task, we select the functions to be
\begin{equation}\label{win}
w_z(x,y)=s_z\ T_{\tilde{x}_1,\tilde{z}_1}T_{\tilde{x}_2,\tilde{z}_2} T_{\tilde{y}_1,\tilde{z}_1}T_{\tilde{y}_2,\tilde{z}_2},
\end{equation}
where $s_z\in\{+1,-1\}$ is a degree of freedom. We write $x=(\tilde{x}_1,\tilde{x}_2)\in[4]^2$, and analogously for $y$ and $z$, and define the matrix $T$ as
\begin{equation}\label{eqT}
	T= \begin{pmatrix}
		1&\phantom{-}1&\phantom{-}1&\phantom{-}1\\
		1&\phantom{-}1&-1&-1\\
		1&-1&\phantom{-}1&-1\\
		1&-1&-1&\phantom{-}1
	\end{pmatrix},
\end{equation}
where we note that $T$ is a so-called Hadamard matrix. The success in performing the task can then naturally be quantified through the average expectation value of outputting the function values,
\begin{equation}\label{witness}
W(D)\equiv \frac{1}{16^3}\sum_{x,y,z}w_z(x,y) \ E_{xyz},
\end{equation} 
where $E_{xyz}=p(c=+1|x,y,z)-p(c=-1|x,y,z)$ is the expectation value of Charlie's outcome. \\

\subsection{Separable correlation bounds}
Consider that Alice and Bob have no access to entanglement, i.e.~that the shared state $\rho_{AB}$ is separable. This means that Alice and Bob are effectively sharing a classical random variable and have locally access to quantum states. We now show that the largest value of $W(D)$ possible under arbitrary encodings for Alice and Bob, and arbitrary measurements for Charlie, is
\begin{equation}\label{cl4}
\!\!\!W_{\text{Sep}}(D)\!\equiv\!\!\max_{\tau_x^A,\tau_y^B, C_z} \!\frac{1}{16^3}\sum_{x,y,z} w_{z}(x,y) \tr\left(\tau^A_x\!\otimes\! \tau^B_y C_z\right)\! \leq \frac{D}{16}.
\end{equation}
Note that due to the linearity of $W(D)$, the optimal value is achieved with a deterministic strategy, corresponding to a (shared) product state $\tau^A \otimes \tau^B$ instead of a general separable state. %

To prove Eq~\eqref{cl4}, we define $f_{xz}=T_{\tilde{x}_1,\tilde{z}_1}T_{\tilde{x}_2,\tilde{z}_2}$, $O^A_{z}=\sum_x f_{xz}\tau^A_x$ and $f_{yz}=T_{\tilde{y}_1,\tilde{z}_1}T_{\tilde{y}_2,\tilde{z}_2}$, $O^B_{z}=\sum_y	 f_{yz}\tau^B_y$. In this notation, the score becomes $W(D)=\frac{1}{16^3} \sum_z s_z \tr\left(O^A_z\otimes O^B_z C_z\right)$. This lets us identify that the structure of the optimal $C_z$ is as follows: both the systems $A$ and $B$ must be projected onto their positive or negative eigenspaces, with a sign determined by the coefficient $s_z$. Hence, Charlie can measure $\tau^A_x$ and $\tau^B_y$ separately and obtain the respective binary outcomes $c_A$ and $c_B$, and then output $c=s_z c_Ac_B $. To this end, we write the spectral decompositions of the operators $O_z^A$ and $O^B_z$ as $O^l_z=\sum_{i=1}^D \mu_{i,z}^l\ketbra{\phi^l_{i,z}}{\phi^l_{i,z}}$ for $l\in\{A,B\}$. Thus, the optimal observable becomes $C_z=s_z\sum_{j,k} \text{sgn}(\mu_{j,z}^A)\text{sgn}(\mu_{k,z}^B) \ketbra*{\phi^A_{j,z}}{\phi^A_{j,z}}\otimes \ketbra*{\phi^B_{k,z}}{\phi^B_{k,z}}$, where $\text{sgn}(\cdot)\in\{+1,-1\}$ denotes the sign of its argument. This leads to  
\begin{align}\nonumber\label{step2}
W_{\text{Sep}}(D) &=\max_{\tau_x^A,\tau_y^B} \frac{1}{16^3}\sum_z \big( \sum_j|\mu_{j,z}^A|\big)\big( \sum_k|\mu_{k,z}^B|\big)\\\nonumber
&\hspace{-1.1cm}\leq \max_{\tau_x^A} \frac{1}{16^3} \sqrt{\sum_z\big( \sum_j|\mu_{j,z}^A|\big)^2}  \max_{\tau_y^B} \sqrt{\sum_z\big( \sum_k|\mu_{k,z}^B|\big)^2}\\
& = \max_{\tau_x} \frac{1}{16^3} \sum_z\big( \sum_j|\mu^A_{j,z}|\big)^2,
\end{align}
where in the second step we have used the Cauchy-Schwarz inequality and in the third step that the two maximisations are independent and identical. We can re-write the right-hand-side in terms of the one-norm; $\sum_z\big( \sum_{j=1}^D|\mu^A_{j,z}|\big)^2 =\sum_z \|O_z\|_1^2$. This is handy because we can then use the operator norm inequality $\|A\|_1\leq \sqrt{\text{dim}(A)}\|A\|_2$, where $\|A\|_2=\sqrt{\tr\left(A^\dagger A\right)}$ and $\|A\|_1=\tr \sqrt{A^\dagger A}$. Applying this term-wise gives  $\sum_z \|O_z\|_1^2\leq D\sum_z \|O_z\|_2^2=D\sum_{x,x'} M_{x,x'}\tr(\tau_x \tau_{x'})$, where we have defined $M_{x,x'}=\sum_zf_{xz}f_{x'z}$. Computing the matrix $M$ from the coefficient matrix \eqref{eqT} reveals that this is merely proportional to the identity matrix, $M_{x,x'}=16\delta_{x,x'}$. Therefore, $W_\text{Sep}(D)\leq \max_{\tau_x} \frac{D}{16^2}\sum_x \tr(\tau_x^2)= \frac{D}{16}$. \qed


It is relevant to ask whether the inequality \eqref{cl4} can be saturated. We first consider the case of $D=4$ since we will later have a particular interest in making this choice. We can then select the preparations of Alice and Bob to be identical, and that each $(D=4)$-dimensional state is composed simply of two independent qubits, namely $\tau_x^A=\tau_x^B=\tau^{(1)}_{\tilde{x}_1}\otimes\tau^{(2)}_{\tilde{x}_2}$. For the set of qubits $\{\tau^{(k)}_{\tilde{x}_k}\}_{\tilde{x}_k}$, for $k\in\{1,2\}$, we select pure states such that they resolve the identity, i.e.~$\sum_{\tilde{x}_k=1}^4\tau^{(k)}_{\tilde{x}_k}=2\openone$. It follows from the Hamadard property of the matrix $T$ in \eqref{eqT}, namely $T\ T^T=4\openone$, that any such set of states saturates the inequality \eqref{cl4}. 
To see this, we write $O_z^A = O_z^B = o^{(1)}_{\tilde{z}_1}\times o^{(2)}_{\tilde{z}_2}$, where $o_{\tilde z_k}^{(k)}=\sum_{\tilde{x}_k=1}^4 T_{\tilde{x_k},\tilde{z}_k}\tau_{\tilde{x}_k}^{(k)}$, $k=1,2$, which leads to $W_\text{Sep}(4)\leq \max_{\{\tau_{\tilde{x}_1},\tau_{\tilde{x}_2}\}} {\frac{1}{4^3}\sum_{\tilde{x}_1}\tr\big[(\tau_{\tilde{x}_1}^{(1)})^2\big]\sum_{\tilde{x}_2}\tr\big[(\tau_{\tilde{x}_2}^{(2)})^2\big]}= 1/4$. This is saturated if and only if $|\lambda_1(o^{(k)}_{\tilde z_k})|=|\lambda_2(o^{(k)}_{\tilde z_k})|$ for each $\tilde{z}_k\in[4]$ and $k\in\{1,2\}$. Here, $|\lambda_{1,2}(M)|$ denotes the absolute values of the two eigenvalues of the $2\times 2$ Hermitian matrix $M$. Equality follows from the properties of matrix $T$. In particular, for $\tilde{z}_k=1$ we have $o^{(k)}_1=2\openone$, whose eigenvalues are $\lambda_{1,2}=2$. On the other hand, for $\tilde{z}_k>1$ we have $\tr(o^{(k)}_{\tilde{z}_k})=0$ for $k\in\{1,2\}$ which implies that the eigenvalues satisfy $\lambda_2=-\lambda_1$. Consequently,  many different quantum states can be considered optimal. In particular also classical states can saturate the bound, for example $\tau^{k}_{1}=\tau^{k}_{2}=\ketbra{0}{0}$ and $\tau^{k}_{3}=\tau^{k}_{4}=\ketbra{1}{1}$. This implies  that there is no performance difference between classical models and quantum models without entanglement.

\begin{figure}
	\centering
	\includegraphics[width=\columnwidth]{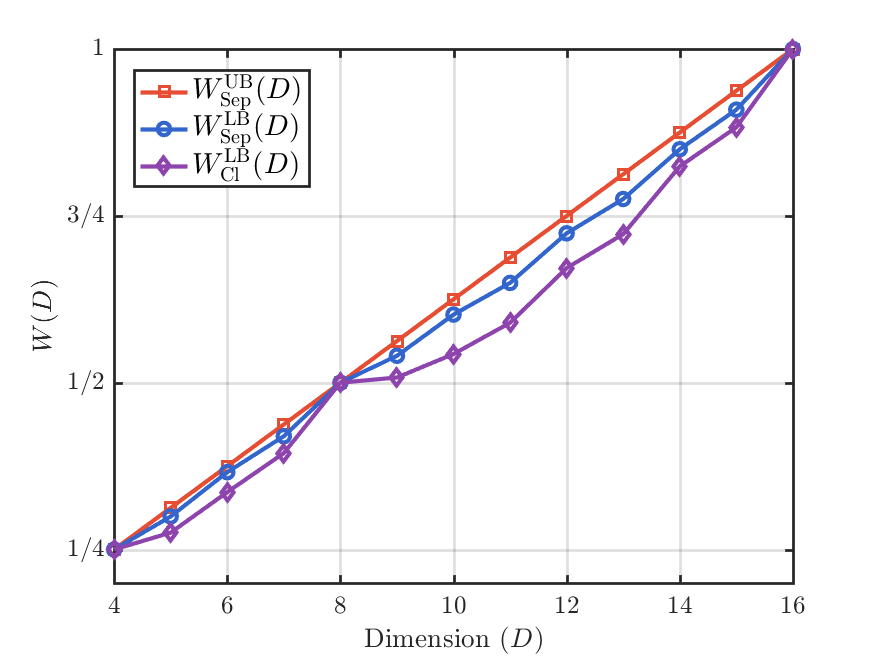}
	\caption{\textit{Heuristic bounds on the witness.} The figure shows lower bounds (LB) and analytical upper bound (UB) for $W_{\text{Sep}}(D)$ across message dimensions $D=4$ to $D=16$. We have considered both sending classical systems (Cl) (states that are diagonal in the computational basis) and quantum systems (Sep) over the channels.}
\label{Fig_WsepD}
\end{figure}

For other values of $D$ the situation is more complicated. We have numerically sought to maximise $W(D)$ up to $D=16$ by using a heuristic search based on alternating convex programs \cite{SDPrev}. We have considered this both for classical messages and for quantum messages; the results are illustrated in Fig.~\ref{Fig_WsepD}. We observe that quantum messages now perform better than their classical counterparts, except for $D=4, 8$ and $D=16$. The results also suggest that our bound \eqref{cl4} is not tight for every $D$.\\

\begin{table*}[!t]
	\centering
	\begin{tabular}{l| c c c c c c c c c c c c c c c c}
		\toprule
		$D$ & 2 & 3 & 4 & 5 & 6 & 7 & 8 & 9 & 10 & 11 & 12 & 13 & 14 & 15 & 16\\
		\midrule
		CCNR (Bloch-diag PPT) & 1.0000 & 1.1786 & 1.5000 & 1.2669 & 1.5000 & 1.3782 & 1.7000 & 1.5556 & 1.6078 & 1.4858 & 1.7679 & 1.5947 & 1.5894 & 1.4932 & 2.2500\\
		CCNR (all PPT) & 1.0000 & 1.1888 & 1.5000 & 1.5000 & 1.5881 &
		1.5881 & 1.7000 & 1.8889 & 1.8889 & 1.8889 & 1.8889 & NaN & NaN & NaN & NaN\\
		\bottomrule
	\end{tabular}
	\caption{\textit{Heuristic bounds for the maximum CCNR-values achievable with PPT states of local dimension $D$.} 
		For Bloch-diagonal states, the CCNR-value is computed over a Hermitian basis formed by the product of normalized Weyl-Heisenberg operators. This is compared with the CCNR-value found when optimising over arbitrary PPT states. In both cases, the optimization was performed using an iterative two-step procedure \cite{Table_2_footnote}.}
	\label{Tab2}
\end{table*}

\subsection{Violation with bound entanglement}
We now show that if Alice and Bob share a state $\rho_{AB}$ that is bound entangled, they can violate the limitation in Eq.~\eqref{cl4}. To this end, we first focus on when the channel capacity is $D=4$ and consider states with the same local dimension. Select the state to be of the Bloch-diagonal form, 
\begin{equation}
\rho_{AB}=\sum_{k=1}^{16} \lambda_{k} G_{k} \otimes G_{k},
\label{belldiag}
\end{equation}
where $\{G_{k}\}$ is an orthonormal basis (i.e.~satisfying $\tr(G_{k} G_{k'})=\delta_{k,k'}$) of four-dimensional Hermitian matrices. To this end, we define $G_{i,j}=\sigma_{i}\otimes\sigma_j$, where $\sigma_{0}, \sigma_{1},\sigma_{2},\sigma_{3}$ are the sub-normalised Pauli operators $\frac{1}{\sqrt{2}}\{\openone,X,Y,Z\}$. A key observation is that by choosing this basis, we have the property that $\tr(G_xG_zG_xG_z)=\tfrac{1}{4}T_{\tilde{x}_1,\tilde{z}_1}T_{\tilde{x}_2,\tilde{z}_2}$ and $\tr(G_yG_zG_yG_z)=\tfrac{1}{4}T_{\tilde{y}_1,\tilde{z}_1}T_{\tilde{y}_2,\tilde{z}_2}$ which pertains to the task conditions in Eq.~\eqref{win}. Due to normalisation we must have $\lambda_1=\frac{1}{4}$ but the remaining $\lambda_z$ can be selected freely as long as $\rho_{AB}\succeq 0$. We choose to relate these coefficients to the variables $s_z$ used in the task conditions \eqref{win} by selecting $s_z=\text{sgn}(\lambda_z)$.

In our protocol, we let Alice and Bob perform unitaries $U_x=2G_x$ and $V_y=2G_y$ on their  respective shares of $\rho_{AB}$ before relaying the four-dimensional quantum messages to Charlie. Then, we let Charlie measure the product observable  $C_z=4G_z\otimes G_z$. The expectation values become  $E_{xyz}=64	\sum_k \lambda_k \tr(G_xG_kG_x^\dagger G_z)\tr(G_y G_k G_y^\dagger G_z)$. A direct calculation shows that $\left|\tr(G_xG_kG_x^\dagger G_z)\right|=\delta_{k,z}/4$, which gives 
\begin{equation}\label{ccnr}
W_\text{BE}(D=4)= \frac{1}{4} \sum_{k}|\lambda_k|= \frac{1}{4} \times \text{CCNR}(\rho_{AB}).
\end{equation}
Here, we have recovered the computable cross-norm or realignment criterion \cite{Chen2003, Rudolph2005}, which states that for all separable states $\text{CCNR}(\rho_{AB})\equiv \|R\|_1\leq 1$, where $R_{kk'}=\tr\left(\rho_{AB}G_k\otimes G_{k'}\right)$. Note that for separable states we obtain exactly the limit $W_\text{Sep}(D=4)$. However, it is well-known that bound entangled states can violate the CCNR-criterion. A relevant example is the state defined by selecting $\lambda_{1}=\frac{1}{4}$ and $|\lambda_z|=\frac{1}{12}$ for $z\neq 1$, with the signs negative for  $z\in\{7,9,11,12,16\}$ and positive otherwise. We call this state $\rho_\text{BE}$. The spectrum of both $\rho_\text{BE}$ and its partial transpose take the form  $\{\left(\frac{1}{6}\right)^{\otimes 6}, 0^{\otimes 10}\}$.  It has a CCNR value of $\text{CCNR}(\rho_\text{BE})=\frac{3}{2}$ and hence generates a significant violation of the limit in Eq.~\eqref{cl4}.

This does not only mean that bound entanglement increases the score in the communication task by $50\%$ but also that these predictions cannot be simulated even if the un-entangled models are permitted five-dimensional messages. The latter follows from Eq.~\eqref{cl4} because  $W_\text{Sep}(D=5)<W_\text{BE}(D=4)$. An alternative way to benchmark the advantage is to consider the mixture of $\rho_\text{BE}$ with white noise of visibility $v\in[0,1]$. The mixed state now becomes $ v\rho_\text{BE}+(1-v)\frac{\mathds{1}}{16}$. The critical visibility needed to break the limit $W(D=4)\leq \frac{1}{4}$ is calculated to $v=\frac{3}{5}$. Thus, the four-dimensional bound entanglement can tolerate $40\%$ white noise before ceasing to violate the inequality.

The above argument leading to Eq.~\eqref{ccnr} points to a connection between the violation of the CCNR-criterion and entanglement-based correlation advantages. We have therefore numerically explored the largest possible CCNR-values reachable with PPT-entangled states. The results are shown in Table~\ref{Tab2} for dimensions up to $D=16$, both for general PPT states and for Bloch-diagonal PPT states. From this, we make a few central observations: (1) For $D=4$ the optimal value found numerically matches that achieved by our state $\rho_{\text{BE}}$. This indicates its optimality. (2) Bloch-diagonal PPT states can sometimes perform just as well as general PPT states. (3) The CCNR-value does not increase monotonically in the dimension $D$ for Bloch-diagonal PPT states.\\

\section{Parallel repetition}
So far, we have found a significant advantage from bound entanglement using four-dimensional systems. Now, we will use this in parallel repetition in order to prove unlimited correlation advantages from bound entanglement in the high-dimensional limit.  To this end, consider that we run the task in $N$ copies. Specifically, Alice and Bob now hold sets of inputs $\vec{x}\equiv x_1\ldots x_N\in[16]^N$ and $\vec{y}\equiv y_1\ldots y_N\in[16]^N$ respectively. They encode quantum states $\tau^A_{\vec{x}}$ and $\tau^B_{\vec{y}}$ of Hilbert space dimension $D$, where $D$ ranges from $1$ to $16^N$. Charlie similarly holds a set of inputs $\vec{z}=z_1\ldots z_N\in[16]^N$ which he associates with a measurement $C_{\vec{z}}$ that has a binary outcome $c\in\{+1,-1\}$. For each $i\in N$, we can associate the functions of interest in the original task, namely $w_{z_i}(x_i,y_i)$. In our $N$-fold parallel repetition of the original task, Charlie's goal is to compute the parity of the function of interest over all values of $i$. The total function thus becomes the product
\begin{equation}
w_{\vec{z}}(\vec{x},\vec{y})=\prod_{i=1}^N w_{z_i}(x_i,y_i).
\end{equation}
The success is again quantified by the average expectation value of outputting the function value, namely  $W^N(D)=\frac{1}{16^{3N}}\sum_{\vec{x},\vec{y},\vec{z}} w_{\vec{z}}(\vec{x},\vec{y})E_{\vec{x}\vec{y}\vec{z}}$, where $E_{\vec{x}\vec{y}\vec{z}}=p(c=+1|\vec{x},\vec{y},\vec{z})-p(c=-1|\vec{x},\vec{y},\vec{z})$ is the expectation value. Note that for $N=1$ this is simply Eq.~\eqref{witness}.

In analogy with Eq.~\eqref{cl4}, we must bound the largest possible value of $W^N(D)$ when Alice and Bob share a separable state, which we label $	W^{N}_\text{Sep}(D)$. In Appendix~\ref{AppProof}, we extend the proof technique from the original task (i.e.~$N=1$) to prove the correlation inequality
\begin{equation}\label{clbound}
	W^{N}_\text{Sep}(D)\leq \frac{D}{16^N}.
\end{equation}

To violate this limit using bound entanglement, we consider a protocol in which Alice and Bob share $N$ copies of our four-dimensional state $\rho_{\text{BE}}$. Thus, the shared state is $\rho_{\text{BE}}^{\otimes N}$. It follows trivially that this state also has PPT. It is also Bloch-diagonal because it can be expressed as
\begin{equation}
\rho_{\text{BE}}^{\otimes N}=\sum_{k_1\ldots k_n}\left(\prod_{l=1}^N \lambda_{k_l}\right) G^N_{\vec{k}}\otimes G^N_{\vec{k}}.
\end{equation}
Here, we write $\vec{k}=(\vec{i},\vec{j})$ with $\vec{i}=i_1\ldots i_N \in[4]^N$ and $\vec{j}=j_1\ldots j_N \in[4]^N$  and define $G^N_{\vec{k}}=G^N_{\vec{i},\vec{j}}\equiv \bigotimes_{l=1}^n\sigma_{i_l}\otimes\sigma_{j_l}$. Note that $\{G^N_{\vec{k}}\}$ is an orthonormal basis of the space of $4^N\times 4^N$ Hermitian matrices.

\begin{figure}[t!]
	\centering
	\includegraphics[width=\columnwidth]{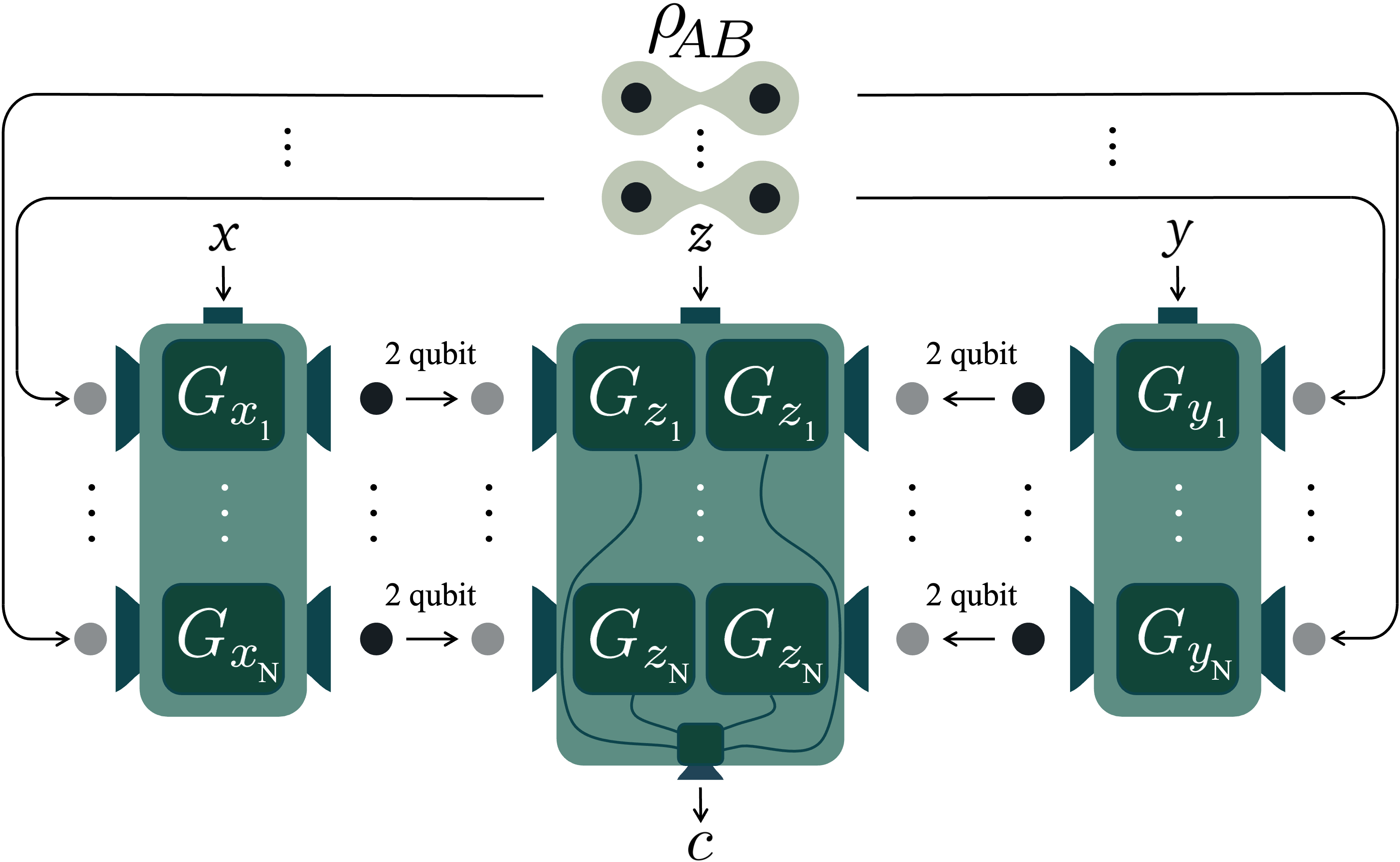}
	\caption{\textit{Parallel repetition using bound entanglement.} Many copies of a two-ququart bound entangled state are shared between Alice and Bob, who perform local rotations on each share. Charlie measures each share individually and classically wires the outcomes to produce the output $c$.}\label{Fig_repetition}
\end{figure}

In the protocol, we let Alice and Bob perform the unitaries $U_{\vec{x}} = 2^NG^N_{\vec{x}}$ and $V_{\vec{y}} = 2^NG^N_{\vec{y}}$ respectively, which correspond to rotating each of the $2N$ qubits in the respective local shares by a Pauli unitary. Then, both the $4^N$-dimensional quantum systems are sent  to Charlie who performs a measurement of the observable $C_{\vec{z}}=4^NG^N_{\vec{z}}\otimes G^N_{\vec{z}}$. Note that this is simply a $4N$-string of standard single-qubit Pauli observables; see Fig~\ref{Fig_repetition}. Evaluating the figure of merit, we find
\begin{align}\label{bep}
W^N_\text{BE}(D=4^N)&=\left(\frac{\text{CCNR}(\rho_{\text{BE}})}{4}\right)^{N}=\frac{1}{4^N}\times\left(\frac{3}{2}\right)^N.
\end{align}
We see that when we set $D=4^N$, corresponding to sending $2N$ qubits per party, the correlations from bound entanglement \eqref{bep} exponentially outperform the limit \eqref{clbound}. This advantage manifests itself in two operationally meaningful way.

Firstly, observe that if we permit a channel capacity of  $D=6^N$ in the model without entanglement, the correlation bound in the inequality \eqref{clbound} equals  the correlations achieved in \eqref{bep} via bound entanglement while using only channel capacity $D=4^N$. In other words, $W^N_\text{BE}(4^N)\geq W^N_\text{Sep}(6^N)$. Hence, no communication overhead smaller than $(\frac{3}{2})^N$ will be sufficient to simulate the predictions of bound entanglement. This implies that the correlations from bound entanglement have a simulation cost exponential in the number of copies.

Secondly, the scalable advantage of bound entanglement is manifest in its noise-tolerance. Consider that the $N$-copy bound entangled state is subject to white noise of visibility $v$. The resulting  mixture becomes $v\rho_{\text{BE}}^{\otimes N}+(1-v)\frac{\mathds{1}}{16^{N}}$. Evaluating the critical visibility for violating the limit \eqref{clbound} yields
\begin{equation}
v^N_\text{crit}=\frac{4^N-1}{6^N-1}.
\end{equation}
For $N=1$ we recover our previous result that $v_\text{crit}^1=\frac{3}{5}=0.60$. Already for two copies, this reduces significantly to $v_\text{crit}^2=\frac{3}{7}\approx 0.43$. In the limit of large $N$, the visibility tends to zero as $O\left(\frac{2^N}{3^N}\right)$. Since the local dimension is $D=4^N$, the scaling of the critical visibility in the Hilbert space dimension is equivalently expressed as $O\left(\frac{1}{D^t}\right)$, where $t=\frac{\ln(3)-\ln(2)}{2\ln(2)}\approx 0.29$. \\

\section{Conclusions}
We have shown that bound entanglement is a scalable resource for quantum correlations that defy arbitrary models that do not feature entanglement as a resource. This advantage can become arbitrarily large by probing the high-dimensional limit of bound entanglement, both in terms of noise-robustness and in terms of communication overhead-cost. Notably, already for four-dimensional systems we observe a sizable correlation advantage which tolerates $40\%$ white noise. This exceeds the largest noise-tolerance known for bound entangled states in Bell inequality tests by orders of magnitude. We have also shown that these types of advantages are not limited to a handful of specific bound entangled states but are  at least as prevalent as the states witnessed through the CCNR-entanglement criterion. This raises the question of whether these high-dimensional bound entangled states  can be used to achieve the first violations of Bell inequalities with significant signal strength.

An interesting feature of our scheme is that it only uses elementary building-blocks. The bound entangled states that we employ in our proof are simply many copies of a four-dimensional bound entangled state. The copies can therefore be generated independently. The encoding operations of Alice and Bob, as well as the decoding operations of Charlie, are not genuinely high-dimensional but act independently on each copy. This means that the protocol does not require  multi-particle entangling operations. Moreover, even the operations on each four-dimensional building-block can be reduced to single-qubit operations since all the encodings and decodings are based on Pauli observables. Protocols with this feature have recently been demonstrated for more standard forms of entanglement \cite{Piveteau2022, zhang2025}. The fact that this simplicity is sufficient to reveal the diverging advantages of bound entanglement suggests a considerable room for improvement in the achievable scaling. More sophisticated schemes that leverage both genuinely high-dimensional encoding operations and entangled measurements for decoding could be able to tap into this potential in a variety of different quantum information tasks. More broadly, our result motivates the open problem of understanding the precise conditions under which bound entanglement ranges from being resourceless to being unlimitedly resourceful in quantum information protocols. \\

\section{Code availability}

The iterative two-step algorithm used to compute the values form Fig.~\ref{Fig_WsepD} is available on GitHub: \url{https://github.com/chalswater/seesaw_Wsep}.

\begin{acknowledgments}
This work is supported by the Swedish Research Council under Contract No.~2023-03498 and the Knut and Alice Wallenberg Foundation through the Wallenberg Center for Quantum Technology (WACQT). LT acknowledges funding from the ANR through the JCJC grant LINKS (ANR-23-CE47-0003).
\end{acknowledgments}

\bibliography{references_BE}

\appendix

\section{Proof of Eq.~\eqref{clbound}}\label{AppProof}
Alice, Bob and Charlie receive the inputs $\vec{x}$, $\vec{y}$ and $\vec{z}$ respectively, for $\vec{x} = x_1\ldots x_N\in[16]^N$ and similarly for $\vec{y}$ and $\vec{z}$. We consider the cases where Alice and Bob share a separable state. A separable state can always be generated by Alice and Charlie in their local laboratories if they share a classical random variable. For every value of this random variable, they can employ a distinct encoding/decoding strategy. However, since our figure of merit is linear in the probabilities, its optimal value must be reached by some deterministic encoding/decoding strategy. Therefore, we can ignore the shared randomness and consider that Alice and Bob prepare states  $\tau^A_{\vec{x}}$ and $\tau^B_{\vec{y}}$ respectively, with local dimension $D^N$. The  expectation value of Charlie's outcome can then be written as 
\begin{align}
p(c|\vec{x},\vec{y},\vec{z}) = \tr\left[\left(\tau^A_{\vec{x}}\otimes\tau^B_{\vec{y}}\right)C_{\vec{z}}\right] \ .
\end{align}
We aim to find the maximum value of the figure of merit, 
\begin{align} \label{eq:Qgeneric}
W = \frac{1}{16^{3N}}\sum_{\vec{x},\vec{y},\vec{z}} w_{\vec{z}}(\vec{x},\vec{y}) \ \tr\left[\left(\tau^A_{\vec{x}}\otimes\tau^B_{\vec{y}}\right)C_{\vec{z}}\right] \ ,
\end{align}
where we have defined $w_{\vec{z}}(\vec{x},\vec{y}) = \prod_{i=1}^N w_{z_i}(x_i,y_i)$ and
\begin{align}
w_{z_i}(x_i,y_i) = s_{z_i}T_{\tilde{x}_{1i},\tilde{z}_{1i}}T_{\tilde{x}_{2i},\tilde{z}_{2i}}T_{\tilde{y}_{1i},\tilde{z}_{1i}}T_{\tilde{y}_{2i},\tilde{z}_{2i}} \ ,
\end{align}
where $s_{z_i}\in\{+1,-1\}$. We write $x_i=(\tilde{x}_{1i},\tilde{x}_{2i})\in[4]^{2}$, and similarly for $y_i$ and $z_i$. The matrix $T$ is defined as
\begin{align}
T=\begin{pmatrix}
1 & \phantom{-}1 & \phantom{-}1 & \phantom{-}1 \\
1 & \phantom{-}1 & -1 & -1 \\
1 & -1 & \phantom{-}1 & -1 \\
1 & -1 & -1 & \phantom{-}1
\end{pmatrix} \ .
\end{align}
To this end, we define the coefficients
\begin{align}
f_{x_i,z_i}=T_{\tilde{x}_{1i},\tilde{z}_{1i}}T_{\tilde{x}_{2i},\tilde{z}_{2i}}
\end{align} 
and the operators 
\begin{align}
O^{A}_{\vec{z}} \!=\! \frac{1}{16^N}\sum_{\vec{x}}\prod_{i=1}^{N} f_{x_i,z_i}\tau^{A}_{\vec{x}} , \ Q^{B}_{\vec{z}} \!=\! \frac{1}{16^N}\sum_{\vec{y}}\prod_{i=1}^{N} f_{y_i,z_i}\tau^{B}_{\vec{y}} .
\end{align}
We then re-write the figure of merit in \eqref{eq:Qgeneric} as
\begin{align}
W= \frac{1}{16^{N}}\sum_{\vec{z}} s_{\vec{z}} \ \tr\left[\left(O^A_{\vec{z}}\otimes O^B_{\vec{z}}\right)C_{\vec{z}}\right] \ ,
\end{align}
for $s_{\vec{z}}=\prod_{i=1}^{N}s_{z_i}$. Let us perform the spectral decomposition of these new operators, 
\begin{align}
O^{A}_{\vec{z}} = \sum_{j=1}^{D} \mu_{j,\vec{z}}^{A}\ket{\phi_{j,\vec{z}}^{A}}\bra{\phi_{j,\vec{z}}^{A}} \ , 
\end{align}
and similarly for $O^{B}_{\vec{z}}$. It follows that the optimal choice of $C_{\vec{z}}$ to maximize $W$ will be the projector onto the positive eigenspace of $O^{A}_{\vec{z}} \otimes O^{B}_{\vec{z}}$. That is,
\begin{align}
C_{\vec{z}} = s_z\sum_{j,k}{\rm sgn}(\mu_{j,\vec{z}}^{A}){\rm sgn}(\mu_{k,\vec{z}}^{B}) \ket{\phi_{j,\vec{z}}^{A}}\bra{\phi_{j,\vec{z}}^{A}} \otimes \ket{\phi_{k,\vec{z}}^{B}}\bra{\phi_{k,\vec{z}}^{B}} \ ,
\end{align}
where ${\rm sgn}(\cdot)\in\{+1,-1\}$ is the sign function. This yields the following bound on the figure of merit
\begin{align}
W_{\rm sep} \leq& \underset{\tau^{A}_{\vec{x}},\tau^{B}_{\vec{y}}}{\max}\frac{1}{16^N}\sum_{\vec{z}} \left(\sum_{j=1}^{D} \left|\mu_{j,\vec{z}}^{A}\right| \right)\left(\sum_{k=1}^{D} \left|\mu_{k,\vec{z}}^{B}\right|\right) \nonumber \\
\leq& \underset{\tau^{A}_{\vec{x}},\tau^{B}_{\vec{y}}}{\max}\frac{1}{16^N}\sqrt{\sum_{\vec{z}} \left(\sum_{j=1}^{D} \left|\mu_{j,\vec{z}}^{A}\right| \right)^2}\sqrt{\sum_{\vec{z}} \left(\sum_{k=1}^{D} \left|\mu_{k,\vec{z}}^{B}\right|\right)^2} \nonumber \\
=& \underset{\tau^{A}_{\vec{x}}}{\max}\frac{1}{16^N}\sum_{\vec{z}} \left(\sum_{j=1}^{D} \left|\mu^A_{j,\vec{z}}\right| \right)^2 \ ,
\end{align}
where in the second step we used the Cauchy-Schwarz inequality, and in the last step the fact that the optimisation is reduced to two independent maximizations. Let us now make use of the norm inequality
\begin{align} \label{eq:opnorm}
||O_{\vec{z}}||_1 \leq \sqrt{D}||O_{\vec{z}}||_2 \ ,
\end{align}
where
\begin{align}
||O_{\vec{z}}||_1 = \sum_{j=1}^{D}|\mu_{j,\vec{z}}| \ , \quad  ||O_{\vec{z}}||_2 = \sqrt{\tr\left((O_{\vec{z}})^2\right)} \ .
\end{align}
Replacing \eqref{eq:opnorm} and squaring both sides one gets
\begin{align}
\frac{1}{16^N}\sum_{\vec{z}}\left(\sum_{j=1}^{D} \left|\mu^A_{j,\vec{z}}\right| \right)^2 \leq \frac{D}{16^N}\sum_{\vec{z}} \tr\left((O_{\vec{z}})^2\right) \ .
\end{align}
The last term turns into
\begin{align}
& \frac{D}{16^N}\sum_{\vec{z}} \tr\left((O_{\vec{z}})^2\right) \\
= & \frac{D}{16^{3N}} \sum_{\vec{x},\vec{x}'} \left(\sum_{\vec{z}}\prod_{i,j=1}^{N} f_{x_i,z_i} f_{x'_j,z_j}\right)\tr\left(\tau_{\vec{x}}\tau_{\vec{x}'}\right) \nonumber \\
=& \frac{D}{16^{3N}} \sum_{\vec{x},\vec{x}'} M_{\vec{x},\vec{x}'} \tr\left(\tau_{\vec{x}}\tau_{\vec{x}'}\right) \leq \frac{D}{16^N} \ , \nonumber
\end{align}
where $M_{\vec{x},\vec{x}'} =\sum_{\vec{z}}\prod_{i,j=1}^{N} f_{x_i,z_i} f_{x'_j,z_j}= 16^{N}\delta_{\vec{x},\vec{x}'}$, and the sum over $\vec{x}$ runs over $16^N$ elements. This means that,
\begin{align}
W_{\rm Sep} \leq \frac{D}{16^N}  \ .
\end{align}

\end{document}